\documentclass[nofootinbib,
 onecolumn,
 amsmath,amssymb,
 aps,
 prd,
 showpacs,
 superscriptaddress,
]{revtex4-1}

\usepackage{mathrsfs}
\usepackage{xspace}
\usepackage{gensymb}
\usepackage{comment}
\usepackage{graphicx}
\usepackage{dcolumn}
\usepackage{bm}
\usepackage{color}
\usepackage{xcolor}

\usepackage[normalem]{ ulem }
\usepackage{soul}

\usepackage[utf8]{inputenc}

\usepackage{accents}
\newlength{\dhatheight}
\newcommand{\doublehat}[1]{
    \settoheight{\dhatheight}{\ensuremath{\hat{#1}}}
    \addtolength{\dhatheight}{-0.35ex}
    \hat{\vphantom{\rule{1pt}{\dhatheight}}
    \smash{\hat{#1}}}}

\def\J {\ensuremath{J}\xspace}

\newcommand\CellTopTwo{\rule{0pt}{2.8ex}}

\begin{document}

\title{Search for dark matter annihilation in the dwarf irregular galaxy WLM with H.E.S.S.}

\author{H.E.S.S. Collaboration}
\noaffiliation

\author{H.~Abdallah} 
\affiliation{Centre for Space Research, North-West University, Potchefstroom 2520, South Africa}

\author{R.~Adam}
\affiliation{Laboratoire Leprince-Ringuet, École Polytechnique, CNRS, Institut Polytechnique de Paris, F-91128 Palaiseau, France}

\author{F.~Aharonian}
\affiliation{Max-Planck-Institut f\"ur Kernphysik, P.O. Box 103980, D 69029 Heidelberg, Germany}
\affiliation{Dublin Institute for Advanced Studies, 31 Fitzwilliam Place, Dublin 2, Ireland}
\affiliation{High Energy Astrophysics Laboratory, RAU, 123 Hovsep Emin St Yerevan 0051, Armenia}

\author{F.~Ait Benkhali}
\affiliation{Max-Planck-Institut f\"ur Kernphysik, P.O. Box 103980, D 69029 Heidelberg, Germany}

\author{E.O.~Ang{\"u}ner}
\affiliation{Aix Marseille Universit\'e, CNRS/IN2P3, CPPM, Marseille, France}

\author{C.~Arcaro}
\affiliation{Centre for Space Research, North-West University, Potchefstroom 2520, South Africa}

\author{C.~Armand}
\email[]{Corresponding authors. \\  contact.hess@hess-experiment.eu}
\affiliation{Laboratoire d'Annecy de Physique des Particules, Univ. Grenoble Alpes, Univ. Savoie Mont Blanc, CNRS, LAPP, 74000 Annecy, France}

\author{T.~Armstrong}
\affiliation{University of Oxford, Department of Physics, Denys Wilkinson Building, Keble Road, Oxford OX1 3RH, UK}

\author{H.~Ashkar}
\affiliation{IRFU, CEA, Universit\'e Paris-Saclay, F-91191 Gif-sur-Yvette, France}

% ...

\begin{abstract}
We search for an indirect signal of dark matter through very high-energy $\gamma$ rays from the Wolf-Lundmark-Melotte (WLM) dwarf irregular galaxy. The pair annihilation of dark matter particles would produce Standard Model particles in the final state such as $\gamma$ rays, which might be detected by ground-based Cherenkov telescopes. 
Dwarf irregular galaxies represent promising targets as they are dark matter dominated objects with well measured kinematics and small uncertainties on their dark matter distribution profiles.
In 2018, the H.E.S.S. five-telescope array observed the dwarf irregular galaxy WLM for 18 hours. 
We present the first analysis based on data obtained from an imaging atmospheric Cherenkov telescope for this subclass of dwarf galaxy. As we do not observe any significant excess in the direction of WLM, we interpret the result in terms of constraints on the velocity-weighted cross section for dark matter pair annihilation $\langle\sigma v\rangle$ as a function of the dark matter particle mass for various continuum channels as well as the prompt $\gamma \gamma$ emission. For the $\tau^+ \tau^-$ channel the limits reach a $\langle \sigma v \rangle$ value of about $4 \times 10^{-22} \, \text{cm}^3 \text{s}^{-1}$ for a dark matter particle mass of 1~TeV. For the prompt $\gamma \gamma$ channel, the upper limit reaches a $\langle \sigma v \rangle$ value of about $5 \times 10^{-24} \, \text{cm}^3 \text{s}^{-1}$ for a mass of 370~GeV. These limits represent an improvement of up to a factor 200 with respect to previous results for the dwarf irregular galaxies for TeV dark matter search.
\end{abstract}


\maketitle

%%%%%%%%%%%%%%%%%%%%%%%%%%%%%%%%%%%%%%%%%%%%%%%%%
\section{\label{sec:Introduction}Introduction}
%%%%%%%%%%%%%%%%%%%%%%%%%%%%%%%%%%%%%%%%%%%%%%%%%

Astrophysical observations suggest that non-baryonic cold dark matter (DM) represents about 85\% of the matter density in the Universe, affecting the formation of large scale structures, influencing the motion of galaxies and clusters, and bending the path of light. Yet, we do not know much about its nature and properties. 
In the WIMP (Weakly Interacting Massive Particles) paradigm, DM particles may be present in large quantites in dense regions such as dwarf galaxies or the Galactic center. The pair annihilation of DM particles would send indirect signals by creating Standard Model particles, which might be detected \cite{Feng:2010gw,Hooper:2009zm, MultiMessengers}. Among these particles used as probes for indirect DM searches are $\gamma$ rays.
High-energy $\gamma$ rays are not deflected by the magnetic field, so that their source can be well localized in the sky. We use the H.E.S.S. telescopes (High Energy Stereoscopic System) to search for signal of indirect DM annihilation. 

This study focuses on dwarf irregular galaxies, which constitute very promising targets for indirect DM searches~\cite{Gammaldi:2017mio}. 
Dwarf irregular galaxies of the Local Group are located at a distance of a few Mpc, and are mostly rotationally supported with negligible random motion of the gas. These objects are therefore believed to have simple structures and kinematics~\cite{Leaman_2009, R16}.
They represent the smallest stellar systems with extended neutral hydrogen (HI) distributions~\cite{Iorio}. 
This large amount of gas is easily detectable by radio telescopes and is used as a kinematic tracer for deriving the rotation curves up to large radii of the galaxies~\cite{Ghosh_2018}.
While these galaxies have an irregular shape in optical light, they often appear much more regular and symmetrical in radio observations of neutral hydrogen.
The study of these well-constrained rotation curves implies that dwarf irregular galaxies are DM dominated systems~\cite{Oh:2015xoa}. 
These objects are estimated to have high \J factors, a measure of the expected signal from DM annihilations occurring within these sources.
At distances within the Local Group, dwarf irregular galaxies have a DM halo typically extending from $0.3\degree$ to a few degrees in angular radius~\cite{Gammaldi:2017mio}.

The galaxy called Wolf-Lundmark-Melotte (WLM) is one of the most promising dwarf irregular galaxies for a DM search, since it offers one of the largest \J factors among this type of object~\cite{Gammaldi:2017mio}. 
In 2018, the H.E.S.S. experiment observed WLM, which makes H.E.S.S. the first imaging air Cherenkov telescopes to present an analysis on this class of potential $\gamma$-ray source. 
In this article, we present an analysis of the $\gamma$-ray events observed in the direction of the irregular galaxy WLM. 
This new search complements the studies carried out for DM from dwarf spheroidal galaxies~\cite{Abramowski:2014tra, Abdalla:2018mve, Abdallah:2020sas} and from the Galactic center~\cite{Abdallah:2016ygi, Abdallah:2018qtu}. 
Compared to dwarf irregular galaxies, dwarf spheroidal galaxies have larger \J factors, but the knowledge on \J is limited by the number of spectroscopic measurements of individual stars. In comparison, the dwarf irregular galaxies benefit from the numerous measurements of the gas tracer, which allows smaller uncertainties on their \J factors. The HAWC experiment put limits on the DM annihilation cross section for a set of irregular galaxies~\cite{Cadena:2017ldx}. In Ref.~\cite{Gammaldi:2017mio}, the authors used Fermi-LAT data to show sensitivity limits for seven dwarf irregular galaxies.

This article is organized as follows: in Sec.~\ref{sec:pheno} we present the properties of the dwarf irregular galaxy WLM, we recall the $\gamma$-ray flux computation as well as the \J factor calculation, and we describe a DM density profile called \textit{coreNFW} accounting for the star-formation history. We derive the \J factor of WLM using this density profile and show its impact on the uncertainty of \J. In Sec.~\ref{sec:Results}, we show the results of the data analysis from 18 hours of observations, and we present the method of the log-likelihood ratio test statistic in order to derive upper limits on the annihilation cross-section $\langle \sigma v\rangle$ for several annihilation channels and various DM masses. We later discuss the results obtained on $\langle \sigma v\rangle$ and conclude on this work in Sec.~\ref{sec:Conclusion}.

%%%%%%%%%%%%%%%%%%%%%%%%%%%%%%%%%%%%%%%%%%%%%%%%%%%%%%%%%%%%%%%%%%%%%%%
\section{\label{sec:pheno}DM induced $\gamma$-ray flux and \J factor}
%%%%%%%%%%%%%%%%%%%%%%%%%%%%%%%%%%%%%%%%%%%%%%%%%%%%%%%%%%%%%%%%%%%%%%%

%%%%%%%%%%%%%%%%%%%%%%%%%%%%%%%%%%%%%%%%%%%%%%%%%%%%%%%%%%%%%%%%%%%%%%%
\subsection{\label{sec:General_charac}General characteristics of WLM}
%%%%%%%%%%%%%%%%%%%%%%%%%%%%%%%%%%%%%%%%%%%%%%%%%%%%%%%%%%%%%%%%%%%%%%%

WLM, also known by the names DDO 221, UGCA 444, and LEDA 143, was discovered by Max Wolf in 1909, and identified as a dwarf galaxy by Knut Lundmark and Philibert Jacques Melotte in 1926. This faint galaxy of absolute magnitude -14.7 is located in the constellation of Cetus, $985\pm33$~kpc from the Milky Way at coordinates RA = $00^\mathrm{h}01^\mathrm{m}58^\mathrm{s}$ and Dec = $-15\degree27'39''$ (J2000)~\cite{R18}, corresponding to $l=75.86\degree$ and $b=-73.62\degree$ in Galactic coordinates. The optical size of WLM spans about 2.5~kpc (or $0.15\degree$) at its greatest extent. Its nearest neighbor within the Local Group, the Cetus dwarf spheroidal galaxy, is about 175 kpc away~\cite{Whiting_1999} and, therefore, WLM is believed to have developed independently from the influence of other systems~\cite{REJKUBA}.
WLM hosts a star-forming region at its center~\cite{Tosi, Cadena:2017ldx} where the star formation rate is
about $10^{-3} \, M_\odot \mathrm{yr}^{-1}$, a low value suggesting the dwarf galaxy is in quiescent phase at the present time.
The expected $\gamma$-ray flux associated with this star-forming region is $\sim 10^{-15} \, \mathrm{TeV}\mathrm{cm}^{-2}\mathrm{s}^{-1}$~\cite{Gammaldi:2017mio},
negligible compared to the expected signal from the thermal relic DM at TeV scale.
This dwarf of $50$~kpc halo extension shows a smooth HI distribution with a well-measured photometry and stellar kinematics~\cite{R16, R18}. WLM is viewed as a highly inclined oblate spheroid \cite{Leaman_2012} of $74 \pm 2.3\degree$~\cite{R16}.
In addition, WLM is rotationally supported with no significant non-circular motions in the gas~\cite{R16}. 
A smooth and well-constrained rotation curve can then be derived from measurements of the gas motion out to $\sim 3$~kpc~\cite{Karukes:2016eiz}.
A total dynamical mass of $(8 \pm 2) \times10^9 \, M_{\odot}$~\cite{R18} is obtained compared to a total gas mass of 
$8.0 \times 10^7 \, M_{\odot}$~\cite{Oh:2015xoa} and a total stellar mass of $(1.6\pm 0.4) \times 10^7 \, M_{\odot}$~\cite{Zhang, R18,Oh:2015xoa} 
which implies that WLM is DM dominated since only 1.2\% of the dynamical mass is accounted for by gas and stars.

%%%%%%%%%%%%%%%%%%%%%%%%%%%%%%%%%%%%%%%%%%%%%%%%%%%%%%%%%%%%%%%%%%%%%%%
\subsection{\label{sec:Flux}DM induced $\gamma$-ray flux}
%%%%%%%%%%%%%%%%%%%%%%%%%%%%%%%%%%%%%%%%%%%%%%%%%%%%%%%%%%%%%%%%%%%%%%%
WIMPs are expected to annihilate into Standard Model particles whose interactions via hadronization or decay would produce observable $\gamma$ rays.
The expected differential $\gamma$-ray flux (in $\text{m}^{-2} \text{s}^{-1} \text{GeV}^{-1}$) in a solid angle $\Delta \Omega$ produced by DM annihilation, assuming WIMPs are Majorana particles, reads~\cite{Bergstrom:1997fj}:
\begin{equation}
 \frac{d\Phi_{\gamma}}{dE_{\gamma}}(\Delta \Omega) =\displaystyle{\frac{1}{2} \frac{\langle\sigma v\rangle}{4\pi m_{\chi}^2}} \sum_f B_f  \frac{dN_{\gamma}^f}{dE_{\gamma}} \times \J(\Delta \Omega),
\label{big_formula}
\end{equation}
where $m_{\chi}$ is the DM particle mass, $\langle\sigma v\rangle$ is the annihilation cross section averaged over the velocity distribution, and $dN_{\gamma}^f/dE_{\gamma}$ is the differential spectrum of each annihilation channel $f$ with their branching ratio $B_f$.
The last term of the equation is the so-called astrophysical \J factor which describes the DM distribution in the source and determines the strength of the signal emitted by the DM annihilation. Its expression reads:
\begin{equation}
  \J(\Delta \Omega) = \int_{\Delta \Omega} \int_{s} \: \rho_{\rm{DM}}^2(r(s, d, \theta)) \:ds \: d\Omega.   \label{dwarf_J} \\
\end{equation}

This key component contains the DM density profile $\rho_{\text{DM}}$. It is assumed to be spherically symmetric and hence depends only on the distance $r$ from the center of the galaxy. The radius $r$ can also be expressed in terms of the angular radius $\theta$, related to the solid angle $\Delta \Omega$. The parameter $s$ is the distance along the line of sight towards the considered extension of the source.
The distance $r$ can be expressed as $r^2(s, d, \theta) = s^2 +d^2 -2sd\cos\theta$, where $d$ is the distance to the source. The squared density is integrated over a sphere of radius $r$ associated to the angular radius $\theta$ at which we perform the study. This implies an integral over $\Delta \Omega$ and over the line of sight $s$. The limits of the line of sight are derived by solving the equation of $r(s, d, \theta)$ for $s$ with $r=R_{\rm{vir}}$ where $R_{\rm{vir}}$ is the virial radius, defined in the next section, yielding $s_{\rm{max/min}} = d\cos\theta \pm \sqrt{R^2_{\rm{vir}} - d^2 \sin^2 \theta}$.

%%%%%%%%%%%%%%%%%%%%%%%%%%%%%%%%%%%%%%%%%%%%%%%%%%%%%%%%%%%%%%%%%%%%%%%
\subsection{\label{sec:DM_distribution}DM distribution}
%%%%%%%%%%%%%%%%%%%%%%%%%%%%%%%%%%%%%%%%%%%%%%%%%%%%%%%%%%%%%%%%%%%%%%%

The standard $\Lambda$CDM cosmological model predicts for pure DM structures a halo that follows a ``cuspy'' density profile, such as given by the Navarro-Frenk-White (NFW) parametrization ~\cite{Navarro:1995iw}. However, 
gas-rich dwarf irregular galaxies, such as WLM, have a rotation curve that suggests the density distribution is consistent with a cored profile, such as the Burkert parametrization~\cite{Burkert:1995yz}.
In order to explain this behavior, some authors~\cite{R16, R18, Tollet:2015gqa, Pontzen:2014lma} propose a mechanism which transforms cusp profiles progressively into core profiles. In this mechanism called {\it baryonic feedback}, the final DM profile takes into account the history of the stellar component within the galaxy.
The baryonic feedback occurs in active galaxies that are still forming stars which explode at the end of their lives.
Repeated gravitational perturbations by stellar wind and supernova feedback are thought to smooth out the central DM cusp, lowering the central density.
This effect, \emph{DM heating}, repeats over several cycles of star formation, up to the age of the Universe. A review on this topic is available in Ref.~\cite{Pontzen:2014lma}.

We use a fitting function, introduced by~\cite{R16, R18} and called {\it coreNFW}, which includes this cusp-core transformation. The {\it coreNFW} density profile consists of a mixture between the original NFW profile and a corrective term  that takes into account the effect of the baryonic feedback on the DM density distribution. The parametrization of the density can be derived from the original NFW cumulative mass profile $M_{\text{NFW}}(<r)$~\cite{Navarro:1995iw} multiplied by the function $f^n$ that describes the inner density flattening due to the stellar component of galaxies:
\begin{equation}
M_{\text{coreNFW}}(<r) = M_{\text{NFW}}(<r) f^n(r).
\label{eq:McoreNFW}
\end{equation} 

The original NFW DM cumulative mass profile~\cite{Navarro:1995iw} reads
\begin{equation}
M_{\text{NFW}}(<r) = 4 \pi \rho_s r^3_s \left( \ln \left( \frac{r_s + r}{r_s} \right) - \frac{r}{r_s + r} \right),
\end{equation}
where $\rho_s$ is the scale density and $r_s$ is the scale radius.
These scale parameters are directly related to the concentration parameter $c_{\rm{vir}}$ and the virial mass $M_{\rm{vir}}$. Thus, the use of either one of the two sets is equivalent.
The concentration parameter is proportional to the virial radius $R_{\rm{vir}}$ 
while the virial mass defines the mass enclosed within the virial radius $R_{\rm{vir}}$ of a gravitationally bound system. The virial radius $R_{\rm{vir}}$ is defined as the radius at which the density is equal to the product of the critical density $\rho _{c} = 136.05~M_{\odot} \rm{kpc}^{-3}$ of the Universe at the redshift of the system and an overdensity constant $\Delta_{c} = 200$ \cite{R16, R18}. For the galaxy WLM, the virial radius is $R_{\text{vir}} \simeq 50$ kpc, corresponding to $\theta_{\text{vir}} \simeq 2.89\degree$.

The effect of the baryonic feedback on the DM density profile can be modeled by the function $f^n$ responsible for generating a shallower density profile at radii $r<r_c$. This function reads:
\begin{equation}
f^n (r)= \left[ \tanh\left( \frac{r}{r_c}\right)\right]^n
\end{equation}
where the core radius $r_c$ is related to the half-stellar-mass radius $R_{1/2}$ by the coefficient $\eta$, $r_c = \eta R_{1/2}$, with $R_{1/2}$ defined as the radius from the core center that contains half the total stellar mass of the galaxy. 
The coefficient $n$ controls how shallow the core becomes and is tied to the total star formation time $t_{\rm{SF}}$ and the circular orbit time $t_{\rm{dyn}}$, where the latter is a function of $M_{\rm{vir}}$ and $c_{\rm{vir}}$. The parametrization of $n$ is given by Eqs.~(19) and~(20) of Ref.~\cite{R16}.
For dwarf irregular galaxies, this time is taken as the age of the Universe. The coefficient $n$ can take values between $0<n\leq1$ where $0$ corresponds to no core and $1$ to a complete core. 

The density profile $\rho_{\text{coreNFW}}(r)$ can be extracted from the cumulative mass $M_{\text{coreNFW}}(<r)$ in Eq.~(\ref{eq:McoreNFW}) by taking the derivative:
\begin{equation}
\rho_{\text{coreNFW}}(r) = \displaystyle{f^n(r) \rho_{\mathrm{NFW}}(r) + \frac{n f^{n-1}(r) (1-f^2(r))}{4\pi r^2 r_c} M_{\mathrm{NFW}}(<r)},
\label{coreNFW}
\end{equation}
where $\rho_{\mathrm{NFW}}(r)$ is the NFW density profile~\cite{Navarro:1995iw}.
More details on these expressions can be found in~\cite{R16, R18}.

%%%%%%%%%%%%%%%%%%%%%%%%%%%%%%%%%%%%%%%%%%%%%%%%%%%%%%%%%%%%%%%%%%%%%%%
\subsection{\label{sec:J factor}\J factor of WLM}
%%%%%%%%%%%%%%%%%%%%%%%%%%%%%%%%%%%%%%%%%%%%%%%%%%%%%%%%%%%%%%%%%%%%%%%

\begin{figure}[ht!]
\centering{
\includegraphics[scale=0.5]{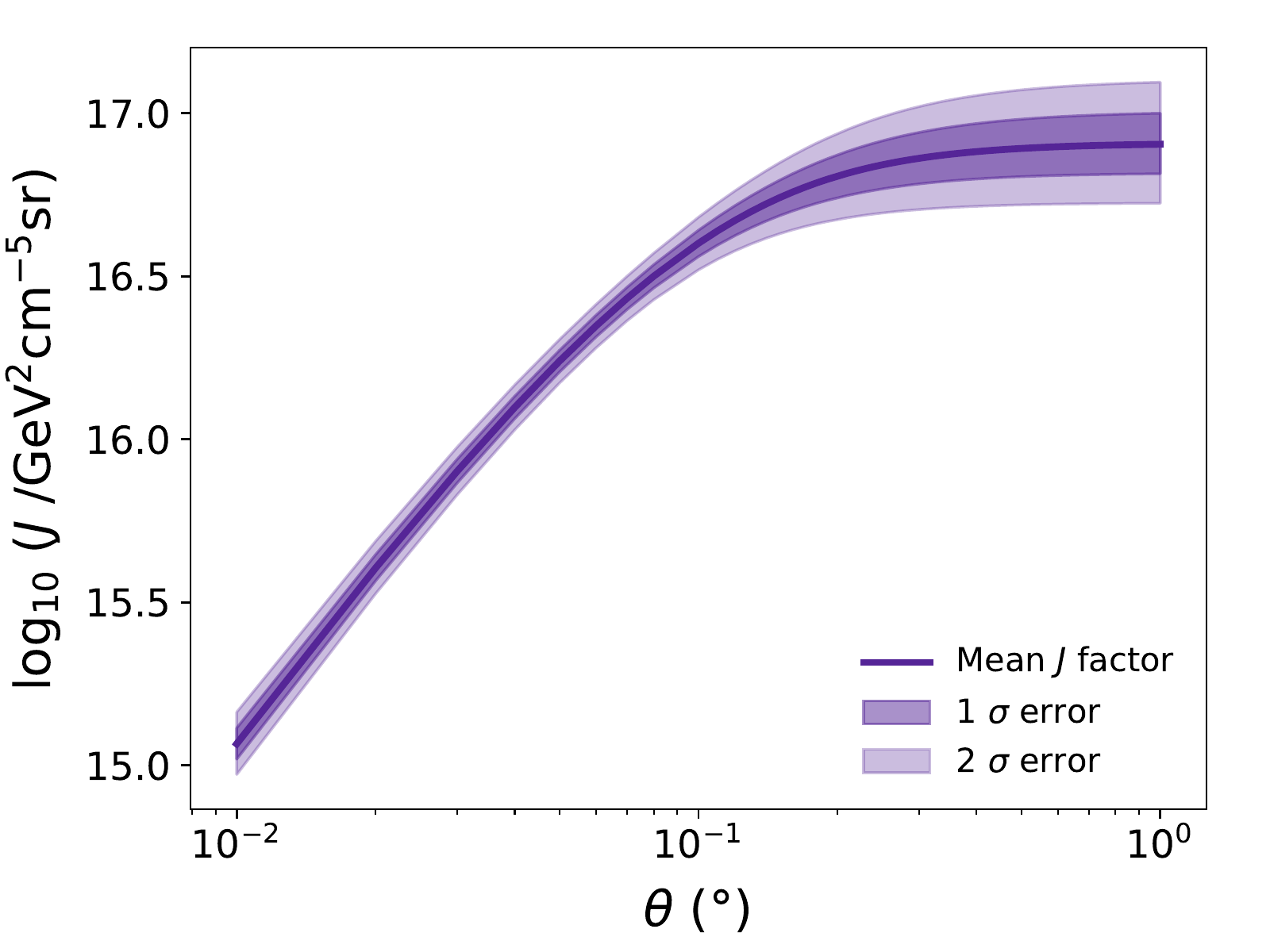}}
\caption{
\J factor as a function of the angular radius $\theta$. The solid line corresponds to the mean values of \J while the dark and light purple bands represent the 1 and 2~$\sigma$ uncertainty bands.}
\label{J_factor_fit}
\end{figure}

To determine the shape of the DM density profile of WLM, one needs to know the parameters associated to the dwarf galaxy in Eq.~(\ref{coreNFW}). 
Following Refs.~\cite{R16, R18, Read}, we use $R_{1/2} = 1.25$ kpc, the projected half-stellar-mass radius of the stars, whose uncertainties are negligible~\cite{Zhang_2012}.
Several measurements of the distance of WLM have been performed ranging from $932$~\cite{McConnachie_2012} to $985$~kpc~\cite{Leaman_2012} with the most likely range being $960 \text{--} 980$~kpc~\cite{Karachentsev:2016cgo,Tully_2013,GildePaz:2006bw}. We use $d = 985 \pm 33$ kpc~\cite{Leaman_2012}, given that the choice of the highest value for the distance leads to a conservative result in the computation of the \J factor.
In Ref.~\cite{R18}, the authors perform a Markov Chain Monte Carlo (MCMC) analysis
based on the spectroscopic data of WLM from the smooth HI distribution, resulting in 
75,000 sets of $M_{\rm{vir}}, c_{\rm{vir}}$, and $\eta$ parameters. 
They use a wide range of priors on these three parameters to explore many different halo configurations that could fit the rotation curve of WLM. The prior ranges are $10^8~M_{\odot} \leq M_{\rm{vir}} \leq 10^{11}~M_{\odot}$, $10.51 \leq c_{\rm{vir}} \leq 21.1$, and $0 < \eta \leq 2.75$~\cite{Read_2017}.
In this article, we use the results of the MCMC analysis from~\cite{Read_2017} and extend their study in order to calculate the \J factor of WLM and its uncertainty~\cite{Read}.
For each of the three-parameter sets ($M_{\rm{vir}}$, $c_{\rm{vir}}$, $\eta$) obtained from the MCMC analysis, we compute the associated parameters $n$, $\rho_s$, $r_s$, and $r_c$, characteristics of the density profile of Eq.~(\ref{coreNFW}), using standard cosmological relations~\cite{R16, R18}.
The average values of their distribution are $n = 0.78$, $\rho_s = 1.53 \times 10^{7}$ $M_\odot \rm{kpc}^{-3}$, $r_s = 3.77$ kpc, and $r_c = 2.49$ kpc. 
We then produce a histogram of \J factors combining Eqs.~(\ref{coreNFW}) and (\ref{dwarf_J}), and using the integration limits as explained in Sec.~\ref{sec:Flux}.
In the computation, we also take into account the uncertainties on the distance of the source. Each of the 75,000 integrals to compute the \J factor is performed with a distance drawn from a Gaussian distribution of mean $d = 985$~kpc and of width $\Delta d = 33$~kpc.
We perform a fit of the distribution with an asymmetric function, where the mean and left/right standard deviations of this fit provide the nominal value and uncertainties of the $\J$ factor.
These quantities are computed for many angular radii $\theta$ of the source which gives the value of the $J$ factor 
shown in Fig.~\ref{J_factor_fit}. 
The \J factor increases, until it reaches a plateau, marking the ``edge'' of the galaxy.
We provide the value of the \J factor, $\log_{10} (\J_{\theta_{\rm{vir}}}  / \text{GeV}^2 \text{cm}^{-5} \rm{sr}) = 16.91 \substack{+0.10\\-0.09}$, for the whole galaxy defined by its virial radius $R_{\rm{vir}}$.
We note that WLM is one of the three best dwarf irregular galaxies for DM search, since only NGC6822 and IC10 have a comparable \J factor on the order of $\log_{10} (\J / \text{GeV}^2 \text{cm}^{-5} \rm{sr}) \simeq 17$~\cite{Gammaldi:2017mio}. We choose to focus on WLM, which has the third largest \J factor among the dwarf irregular galaxies. Regarding IC10, it is located at a position that is not visible by the H.E.S.S. telescopes. As for NGC6822, the galaxy shows a less smooth rotation curve than WLM which would have yielded higher uncertainties on its DM density profile and hence on its \J factor.

%%%%%%%%%%%%%%%%%%%%%%%%%%%%%%%%%%%%%%%%%%%%%%%%%%%%%%%%%%%%%%%%%%%%%%%
\section{\label{sec:Results}Data analysis and results}
%%%%%%%%%%%%%%%%%%%%%%%%%%%%%%%%%%%%%%%%%%%%%%%%%%%%%%%%%%%%%%%%%%%%%%%

%%%%%%%%%%%%%%%%%%%%%%%%%%%%%%%%%%%%%%%%%%%%%%%%%%%%%%%%%%%%%%%%%%%%%%%
\subsection{\label{sec:Observation and data}Observation and dataset}
%%%%%%%%%%%%%%%%%%%%%%%%%%%%%%%%%%%%%%%%%%%%%%%%%%%%%%%%%%%%%%%%%%%%%%%

The H.E.S.S. experiment is an array of five imaging atmospheric Cherenkov telescopes located in central Namibia in the Khomas Highland plateau area at 1,800 meters above sea level. 
The telescopes detect brief flashes of Cherenkov radiation generated by very high energy $\gamma$ rays.
The array consists of four telescopes (CT1-4) with 12 meter reflectors at the corners of a 120 meter square, detecting $\gamma$ rays from $\sim 100$ GeV up to $\sim 100$ TeV. Each reflector is made of spherical mirrors, concentrating the faint Cherenkov flashes on a camera installed in the focal plane of the telescope. The cameras detect and record the signal using an array of photomultipliers. The cameras of the four 12 meter telescopes have been upgraded in 2015/2016, improving their performance and robustness~\cite{HESSIU}.
In 2012, a fifth 28 meter telescope (CT5) was added at the center of the array allowing the event detection down to $\sim 20$ GeV. 

In 2018, H.E.S.S. recorded about 18 hours of good quality data towards WLM with a zenith angle in the range 9 - 51\degree. The energy threshold varies from 120~GeV at low zenith angle to 450~GeV at high zenith angle.
The event reconstruction and classification are performed following standard calibration and selection procedures~\cite{Aharonian:2006pe}. 
We use a technique based on the comparison of the $\gamma$-ray shower image in the camera between real events and the prediction of a semi-analytical model~\cite{Model}. This procedure is based on a log-likelihood minimization using all pixels in the cameras, and provides an optimized use of the five telescope array. We obtain an angular resolution of $0.06\degree$ at 68\% containment radius and a photon energy resolution of 10\% above 200~GeV. The goodness-of-fit parameter and the reconstructed primary depth provided by the semi-analytical model are used to discriminate $\gamma$-ray events from hadronic background.
For the event reconstruction, we use a combined approach~\cite{Holler} which makes use of monoscopic reconstruction (CT5 alone) as well as stereocopic reconstruction (at least two telescopes triggered either for CT1-4 or CT1-5). A $\chi^2$ test allows the determination of the mode that offers the best reconstruction for the $\gamma$-ray-like event. 
Using these selection criteria, an optimal integration radius for a point source is found to be twice the 68\% containment radius of the PSF, or $0.12\degree$.

Since WLM has a DM halo with an angular size that is larger than this region, we investigate if it is better to treat this galaxy as a point-like or extended source in the subsequent analysis. 
We compute the expected signal-to-noise ratio, as a function of the angular radius from the target position, between the expected signal from DM annihilations in WLM and the associated expected background. We find this ratio is maximal at a radius of $0.09\degree$.
As this angle is smaller than the $0.12\degree$ disk used for a point-like source, we treat WLM as a point source in the rest of this article.

Therefore, the region of interest, referred to here as the ``ON region'', is chosen as a disk of radius $0.12\degree$,  
centered at the nominal position of the source. 
The  pointing positions of the telescopes are shifted by $\pm 0.5\degree$ or $\pm 0.8\degree$ with respect to the nominal source position according to the \textit{wobble} mode method~\cite{Berge}. An exclusion region of radius 0.4\degree\ is defined around the ON region to prevent any contamination of the background regions by the tails of the expected DM $\gamma$-ray signal. The background is determined following the multiple-OFF method~\cite{Berge} where 
the OFF region is defined by multiple circular regions of the same size as the ON region and equidistant from the pointing position, \textit{i.e.} the center of the camera.
This method allows the estimation of the residual background and the measurement in the ON region simultaneously so that both are performed under the same observation conditions. The DM signal expected in the OFF regions is estimated to be less than $\sim 1\%$ of the total DM signal. As the ON region and all the OFF regions combined cover a different surface area, the acceptance corrected exposure ratio $\alpha$ is used to renormalize the OFF region surface area to that of the ON region.
Table \ref{table_results_HESS} summarizes the results of the analysis along with the $\gamma$-ray excess and its significance $\sigma$~\cite{Li:1983fv}. We note that the OFF regions differ from a wobble position to another and yield a non integer total $\alpha$.

\begin{figure}[ht!]
\centering{\includegraphics[scale=0.4]{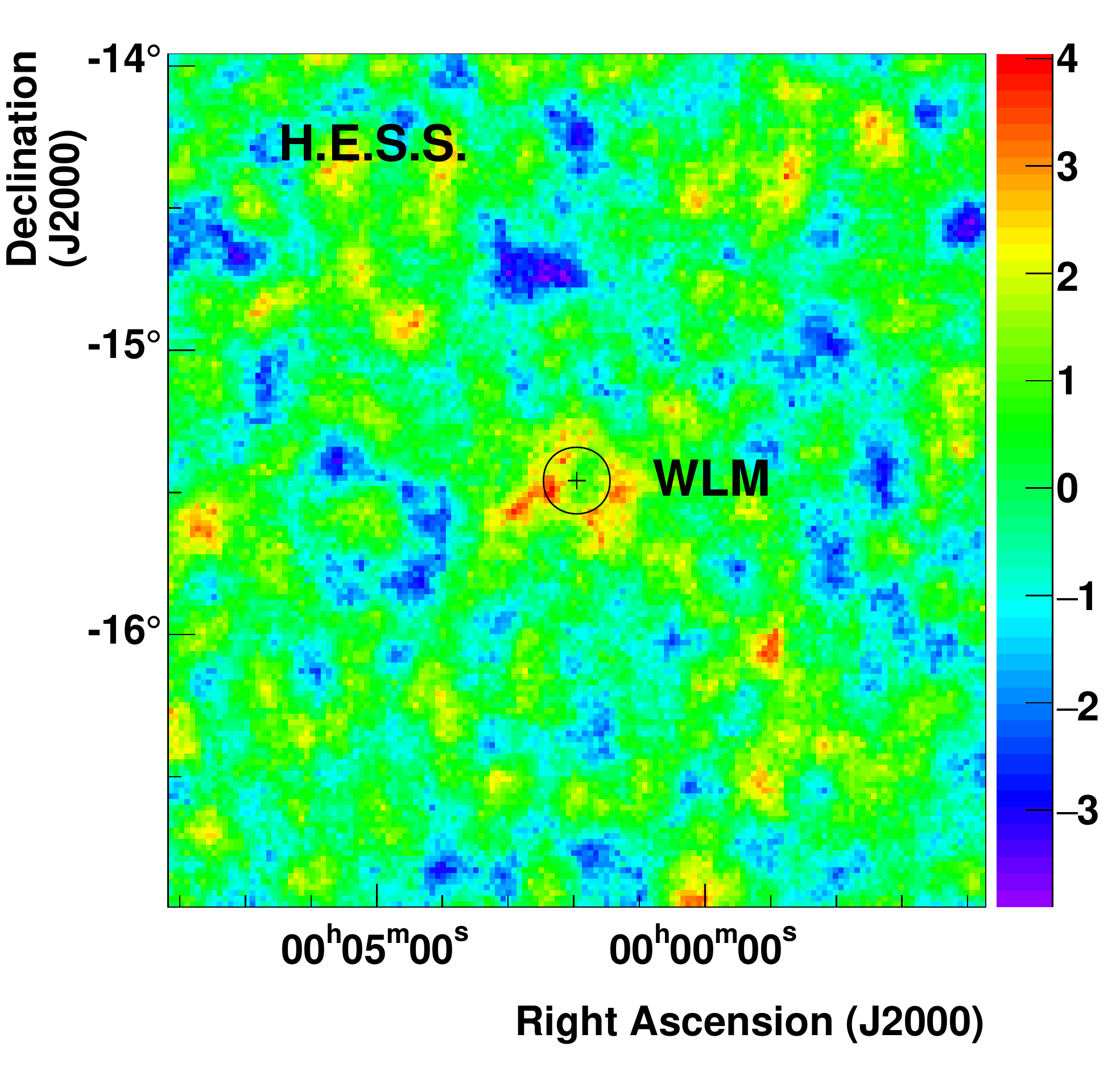}}
\caption{Map of the sky seen by H.E.S.S. in equatorial coordinates showing the significance (color scale) of the $\gamma$-ray excess in number of standard deviations. The black circle indicates the size of the ON region centered around the source position marked with a cross.}
\label{significance}
\end{figure}

\begin{table}[h!]
\small{
\caption{\label{table_results_HESS}Data analysis results of WLM. The quantities $N_{\text{ON}}$ and $N_{\text{OFF}}$ are the number of events detected in the ON and OFF regions, $\alpha$ is the acceptance corrected exposure ratio, the live time correspond to the observation time, $\gamma$ gives the $\gamma$-ray excess detected and $\sigma$ corresponds to the significance of the excess.}
\begin{tabular}{ccccccc}
 \hline
 \hline
 \CellTopTwo{}
~~~~$N_{\text{ON}}$~~~~ & ~~~~$N_{\text{OFF}}$~~~~& ~~~~$\alpha$~~~~ & ~~~$N_{\text{OFF}} / \alpha$~~~ & ~~Live time (hours)~~ & ~~~$\gamma$-ray excess~~~ & ~~~$\sigma$~~~   \\
 \hline
 \CellTopTwo{}
 823  & 11959  & 14.483 & 825.7 & 18.2 & -2.7 & -0.1  \\
 \hline
 \hline
\end{tabular}}
\end{table}
The results presented in Tab. \ref{table_results_HESS} have been cross-checked using a different calibration and analysis chain~\cite{Parsons:2014voa} yielding compatible results.
No significant excess in the signal region is observed in the direction of WLM, or anywhere in the field of view of WLM as can be seen in Fig.~\ref{significance}. Increasing the size of the region of interest up to 0.3\degree\ always leads to significances of the $\gamma$-ray excess below 1~$\sigma$.
Furthermore, the distribution of the significance in the field of view follows a Gaussian function centered on $0$ with a width of $1$, compatible with background fluctuations only.

%%%%%%%%%%%%%%%%%%%%%%%%%%%%%%%%%%%%%%%%%%%%%%%%%%%%%%%%%%%%%%%%%%%%%%%
\subsection{\label{sec:Stats}Statistical analysis}
%%%%%%%%%%%%%%%%%%%%%%%%%%%%%%%%%%%%%%%%%%%%%%%%%%%%%%%%%%%%%%%%%%%%%%%

A log-likelihood ratio test is performed on the data in order to constrain a potential DM signal and set upper limits on the DM annihilation cross section.
We scan over the DM particle mass ranging from 150~GeV to 63~TeV divided into 68 logarithmically-spaced mass bins. The energy bins are also logarithmically-spaced. In order to have enough statistics in each energy bin $i$, a bin containing less than 4 ON or OFF events is merged with the next neighboring bin until the new bin content reaches this threshold~\cite{Feldman:1997qc}.
The total likelihood function $\mathcal{L}$ contains two terms, a product of a Poisson likelihood $\mathcal{L^P}_i$ on the events of all energy bins with a log-normal distribution $\mathcal{L}^J$ of the \J factor. The total likelihood function is written explicitly in terms of the parameter of interest $\langle \sigma v \rangle$ and the nuisance parameters $\boldsymbol{N_\text{B}}$ and \J:
\begin{equation}
\mathcal{L}(\langle \sigma v\rangle, \boldsymbol{N_\text{B}}, \J) = \prod_i \mathcal{L^P}_i (N_{\text{S}_i}(\langle \sigma v\rangle,J), N_{\text{B}_i}|N_{\text{ON}_i}, N_{\text{OFF}_i},\alpha) \times \mathcal{L}^\J(\J|\bar{\J},\sigma_\J).
\end{equation}
For a given energy bin $i$, $N_{\text{S}_i}$ is the number of predicted signal events and $N_{\text{B}_i}$ the number of expected background events with $\boldsymbol{N_\text{B}}$ the corresponding vector.
The values $N_{\text{ON}_i}$ and $N_{\text{OFF}_i}$ are the number of ON and OFF events in the bin $i$, respectively, and $\alpha$ is the acceptance corrected exposure ratio between the ON and OFF regions. 
For an energy bin $i$, the likelihood function $\mathcal{L^P}_i$ of the event counts is the product of two Poisson likelihood functions, one for each of the ON and OFF regions:
\begin{equation}
\mathcal{L^P}_i = \frac{(N_{\text{S}_i}(\langle \sigma v \rangle, J) + N_{\text{B}_i})^{N_{\text{ON}_i}  }}{N_{\text{ON}_i}!} e^{-(N_{\text{S}_i} + N_{\text{B}_i})}
\times \frac{(\alpha N_{\text{B}_i})^{N_{\text{OFF}_i}}}{N_{\text{OFF}_i}!} e^{-\alpha N_{\text{B}_i}}
\end{equation}
where the predicted number of signal events $N_{\text{S}_i}$ in the energy bin $i$ is obtained by performing a convolution of the expected differential $\gamma$-ray flux given by Eq.~(\ref{big_formula}) with the energy resolution function $R(E_{\gamma}, E'_{\gamma})$ relating the energy detected $E'_{\gamma}$ to the true energy $E_{\gamma}$ of the events, the energy-dependent acceptance function $A_{\rm{eff}}(E_{\gamma})$, and the observation time $T_{\rm{obs}}$. The energy resolution is estimated as 10\% over the whole energy range~\cite{Model}. The convolution is then integrated over the bin energy width $\Delta E_i$. The number of signal events for an energy bin $i$ is:
\begin{equation}
 N_{\text{S}_i}(\langle \sigma v \rangle, J) =  J \times \frac{1}{2} \frac{\langle \sigma v \rangle}{4\pi m_{\chi}^2} \int_{\Delta E_i} \int^{\infty}_0   \sum_f B_f \frac{dN^f_{\gamma}}{dE_{\gamma}} \: R(E_{\gamma}, E'_{\gamma}) \: A_{\rm{eff}}(E_{\gamma}) \: T_{\rm{obs}}\: dE_{\gamma} \: dE'_{\gamma}.
\end{equation}

To take into account the uncertainty on the \J factor in our analysis, we introduce in the construction of our total likelihood function $\mathcal{L}$ a log-normal distribution given by
\begin{equation}
\mathcal{L}^\J= \frac{1}{\ln(10) \sqrt{2\pi} \sigma_\J \J} \, \exp{-\frac{( \log_{10}\J - \log_{10}\bar{\J})^2}{2 \sigma^2_\J}},
\label{L^J}
\end{equation}
where $\log_{10}\J$ is the true value of the \J factor and $\log_{10}\bar{\J}$ the value of the mean \J factor with its uncertainty $\sigma_\J$. 
We calculate the \J factor integrated up to $\theta = 0.12\degree$, the size of the ON region, following the procedure detailed in Sec.~\ref{sec:J factor}. We find the value $\log_{10} (\J_{0.12\degree} / \text{GeV}^2 \text{cm}^{-5} \rm{sr}) = 16.68 \pm 0.05$.
We note that this \J factor could have been derived using another DM density profile parametrization, preferably describing a cored DM distribution such as the Burkert profile~\cite{Burkert:1995yz}. Using this profile and the parameters given in Ref.~\cite{Gammaldi:2017mio}, we obtain a relative difference of $-1.7\%$ on $\bar{J}$ within $0.12\degree$ compared to the $\bar{J}$ computed with the \textit{coreNFW} profile.
Assuming that the uncertainties are the same as for the \textit{coreNFW} \J factor, the difference in the upper limits would be negligible.
%However, since no uncertainties are available for the Burkert density profile, we do not derive upper limits using this profile.}

% without which the derivation of upper limits could only be performed with no \J nuisance.

% which would only lead to the derivation of upper limits where no \J nuisance is taken into account.

% which would only lead to the derivation of upper limits without \J as a nuisance.}

We perform a log-likelihood ratio statistical test to set upper limits on the annihilation cross section $\langle \sigma v \rangle$, assuming a positive signal $\langle \sigma v \rangle > 0$, based on the method~\cite{Cowan}: 

\begin{equation}
\rm{TS} = \left\{
\begin{array}{rcl}
&0& \:\: \text{for}  \: \widehat{\langle \sigma v \rangle} > \langle \sigma v \rangle \quad \\ \\
&\displaystyle{-2 \ln \frac{\mathcal{L}(\langle \sigma v \rangle, \doublehat{\boldsymbol{N}}_{\text{B}}(\langle \sigma v \rangle), \doublehat{J}(\langle \sigma v \rangle))}{\mathcal{L}(\widehat{\langle \sigma v \rangle}, \hat{\boldsymbol{N}}_{\text{B}}, \hat{J})}} & \:\: \text{for} \: 0 \leq \widehat{\langle \sigma v \rangle} \leq \langle \sigma v \rangle \quad \\ \\
&\displaystyle{-2 \ln \frac{\mathcal{L}(\langle \sigma v \rangle, \doublehat{\boldsymbol{N}}_{\text{B}}(\langle \sigma v \rangle), \doublehat{J}(\langle \sigma v \rangle))}{\mathcal{L}(0, \doublehat{\boldsymbol{N}}_{\text{B}}(0), \doublehat{J}(0))}} & \:\: \text{for} \: \widehat{\langle \sigma v \rangle} < 0 \quad
\end{array} \right.
\label{coef_qNS0}
\end{equation}
where $\doublehat{\boldsymbol{N}}_{\text{B}}(\langle \sigma v \rangle)$ is the vector of number of background events and $\doublehat{J}(\langle \sigma v \rangle)$ the value of the \J factor maximizing the likelihood function conditionally for a given annihilation cross-section $\langle \sigma v \rangle$. The quantity $\widehat{\langle \sigma v \rangle}$ is the value of the annihilation cross-section, $\hat{\boldsymbol{N}}_{\text{B}}$ the vector of number of background events, and $\hat{J}$ the value of the \J factor that maximize unconditionally the likelihood function. 
In the case of a one-sided test, the criterion value of the test statistic TS is 2.71 corresponding to a 95\% confidence level (CL).
This criterion is used to set the upper limits on the DM velocity-weighted annihilation cross-section $\langle \sigma v \rangle$.

%%%%%%%%%%%%%%%%%%%%%%%%%%%%%%%%%%%%%%%%%%%%%%%%%%%%%%%%%%%%%%%%%%%%%%%
\subsection{Constraints on $\langle \sigma v \rangle$}
%%%%%%%%%%%%%%%%%%%%%%%%%%%%%%%%%%%%%%%%%%%%%%%%%%%%%%%%%%%%%%%%%%%%%%%

\begin{figure}[ht]
 \centering{\includegraphics[scale=0.5]{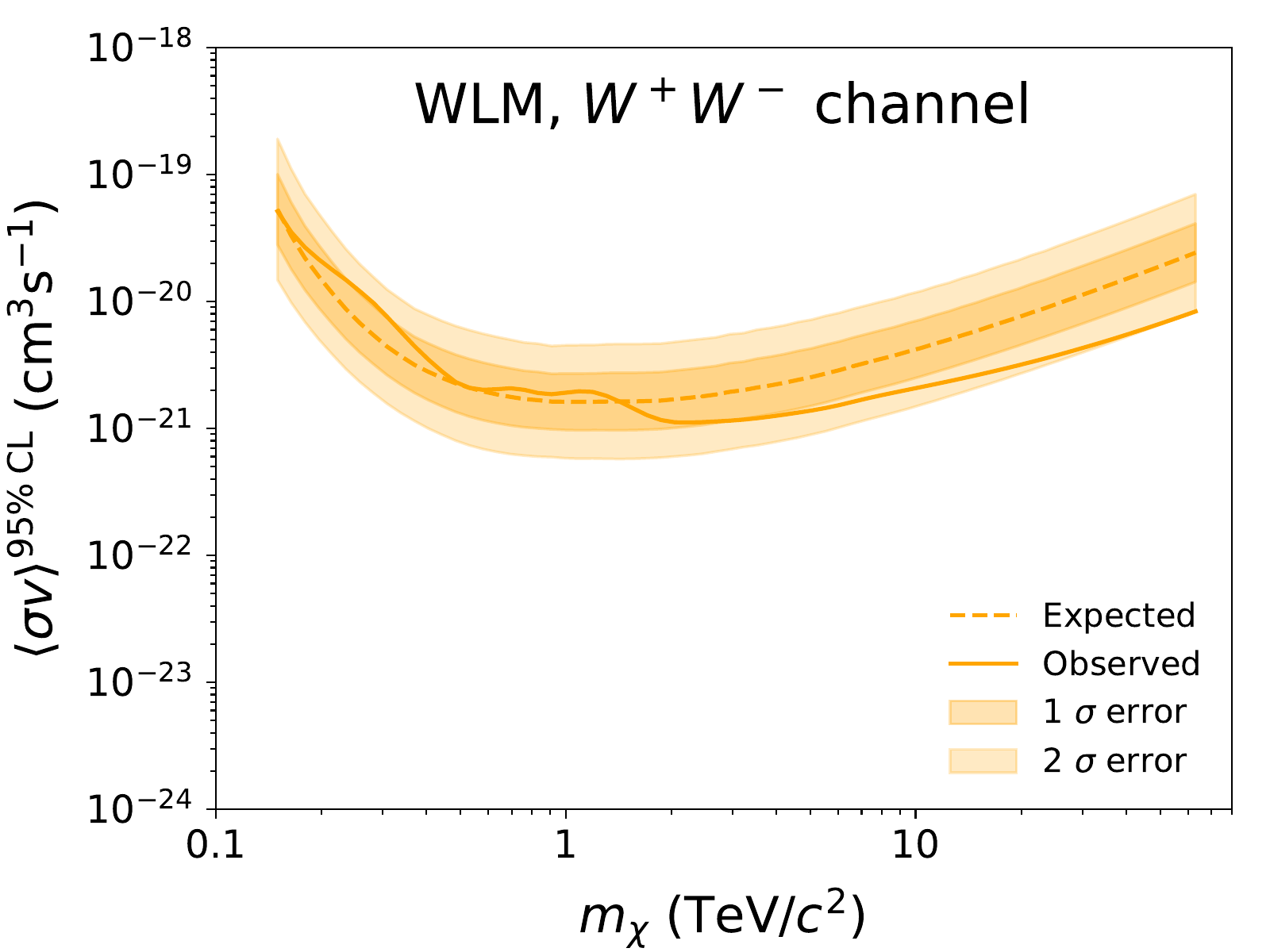} 
 \includegraphics[scale=0.5]{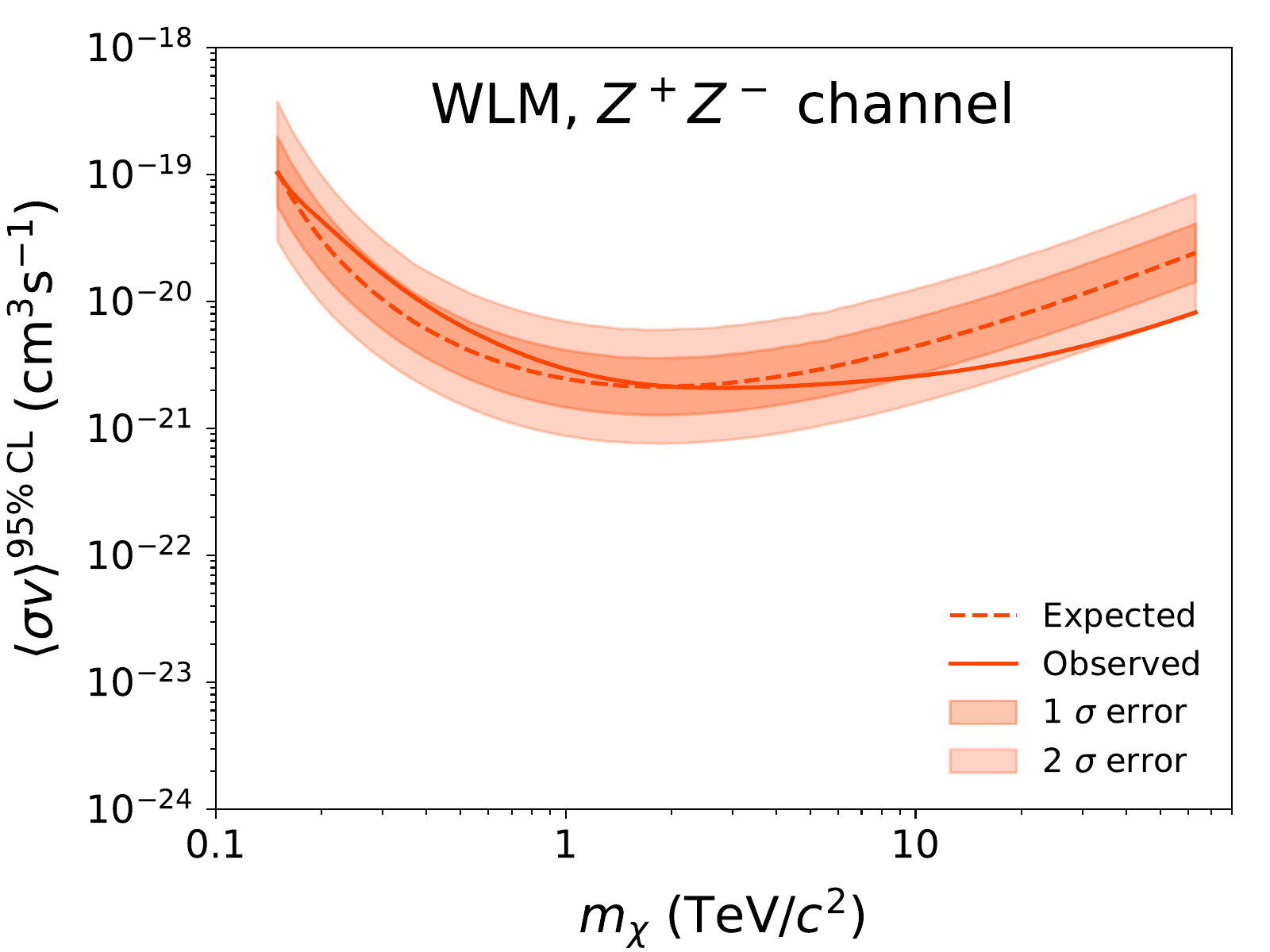}
 \includegraphics[scale=0.5]{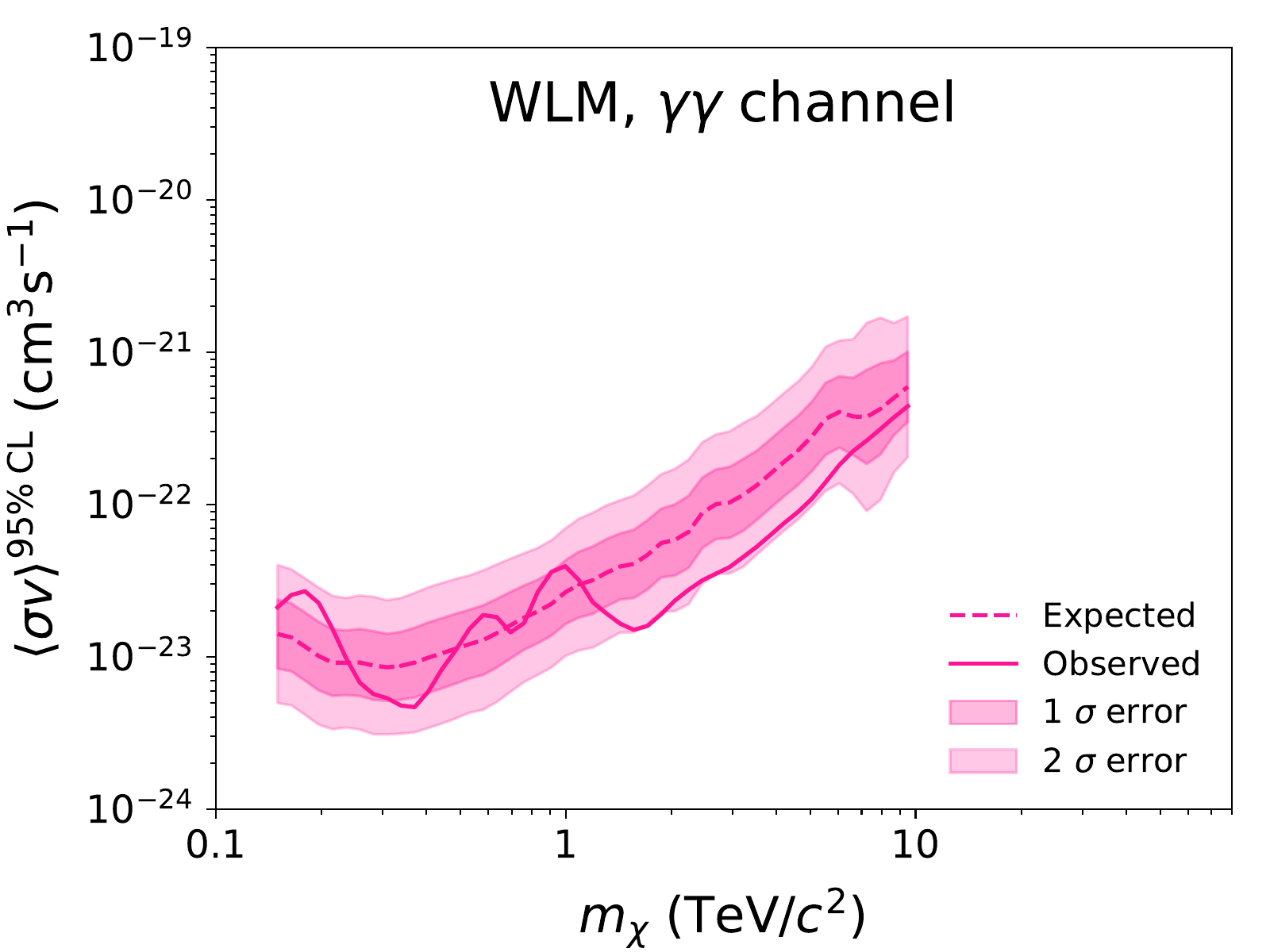} 
 \includegraphics[scale=0.5]{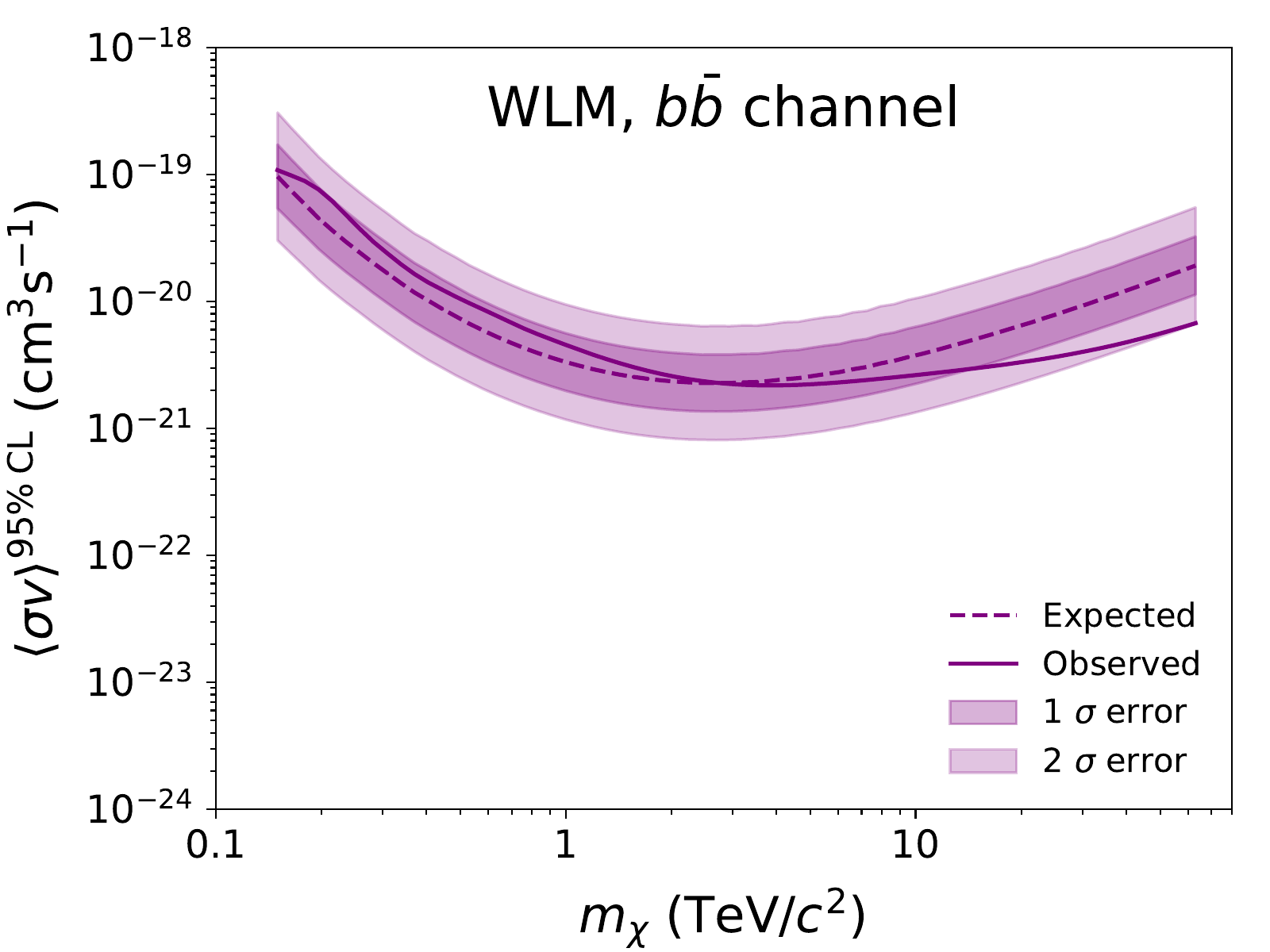}}\\
 \caption{Upper limits on the annihilation cross section $\langle \sigma v \rangle$ at 95\% CL for WLM in the $W^+W^-$,  $Z^+Z^-$, $\gamma \gamma$, and $b\bar{b}$ annihilation channels. These upper limits include the uncertainties on the \J factor. The solid lines are the observed limits, the dashed lines the mean expected limits and the dark (resp. light) bands are the 1~$\sigma$ (resp. 2~$\sigma$) containment bands.}
 \label{UL_sigmav_1}
 \end{figure}

\begin{figure}[ht]
 \centering{\includegraphics[scale=0.5]{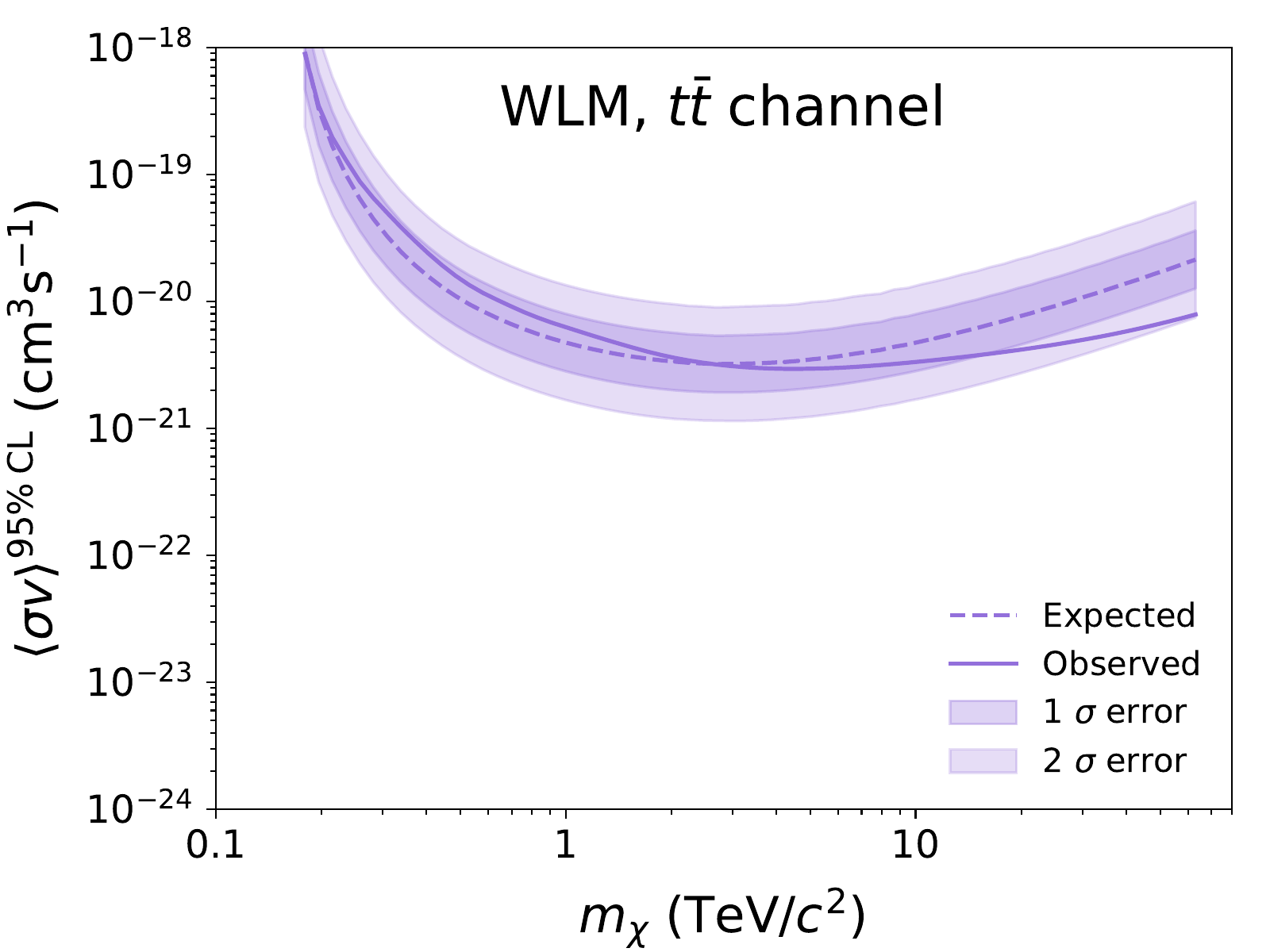} 
 \includegraphics[scale=0.5]{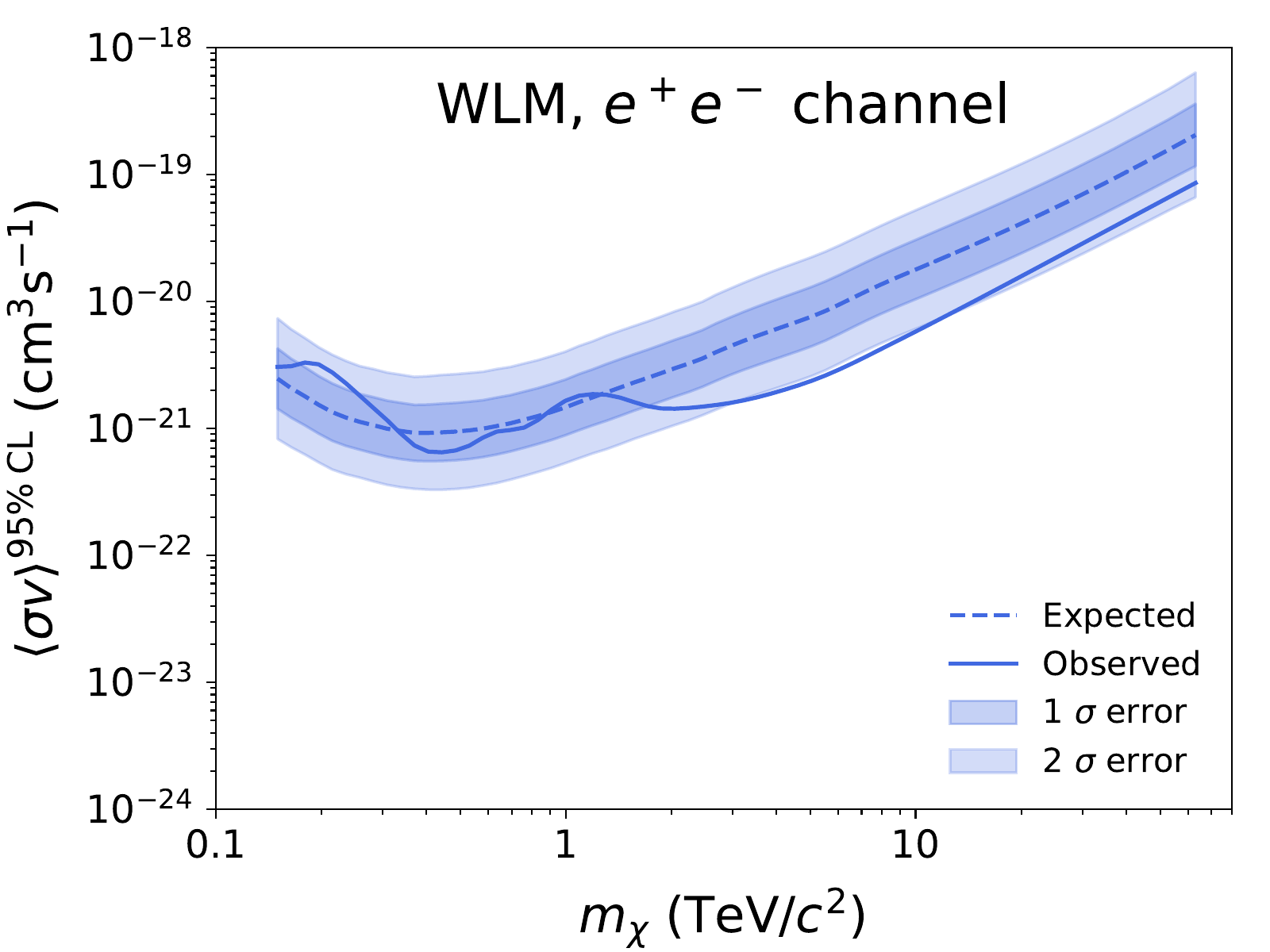}
 \includegraphics[scale=0.5]{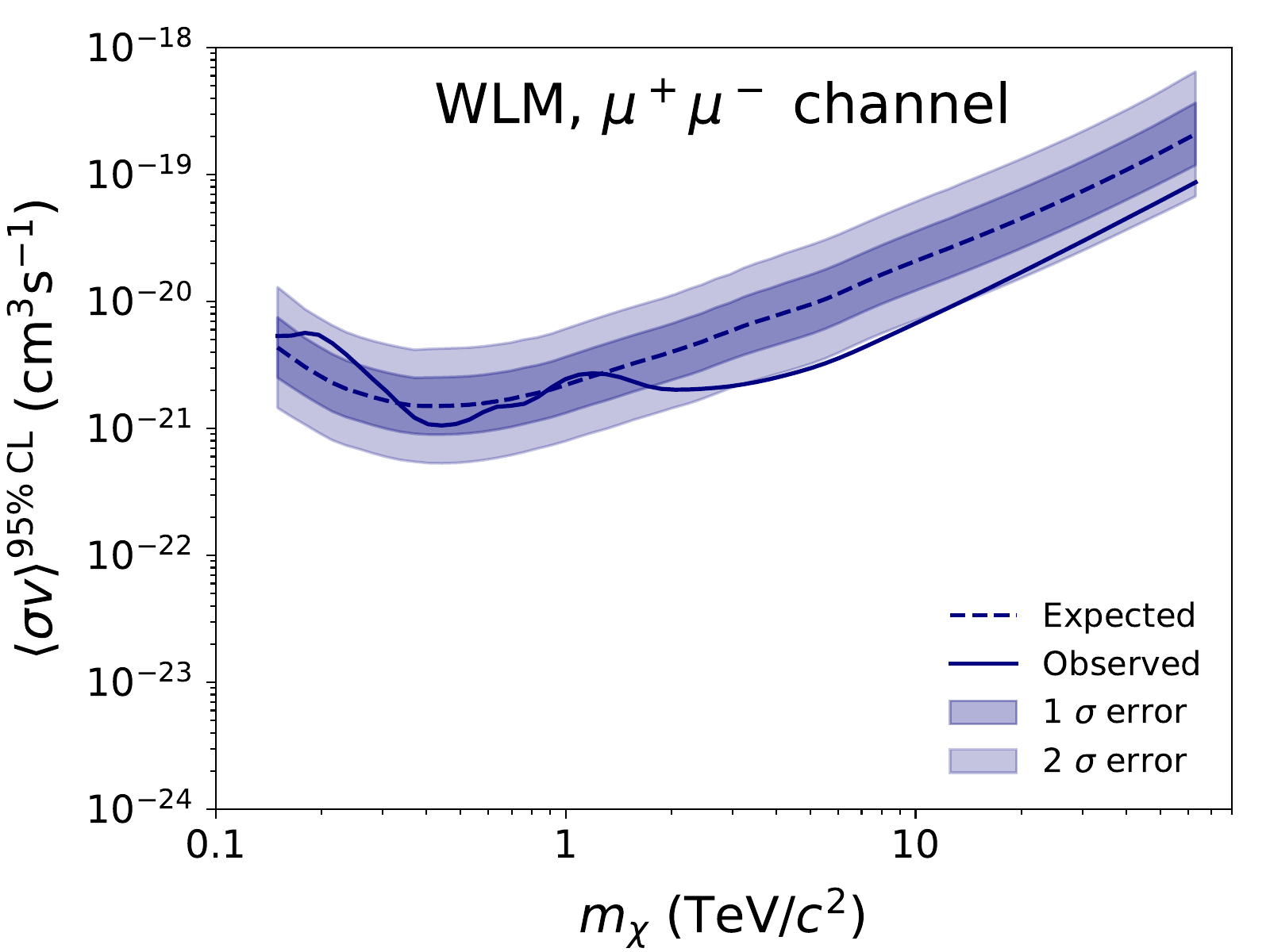} 
 \includegraphics[scale=0.5]{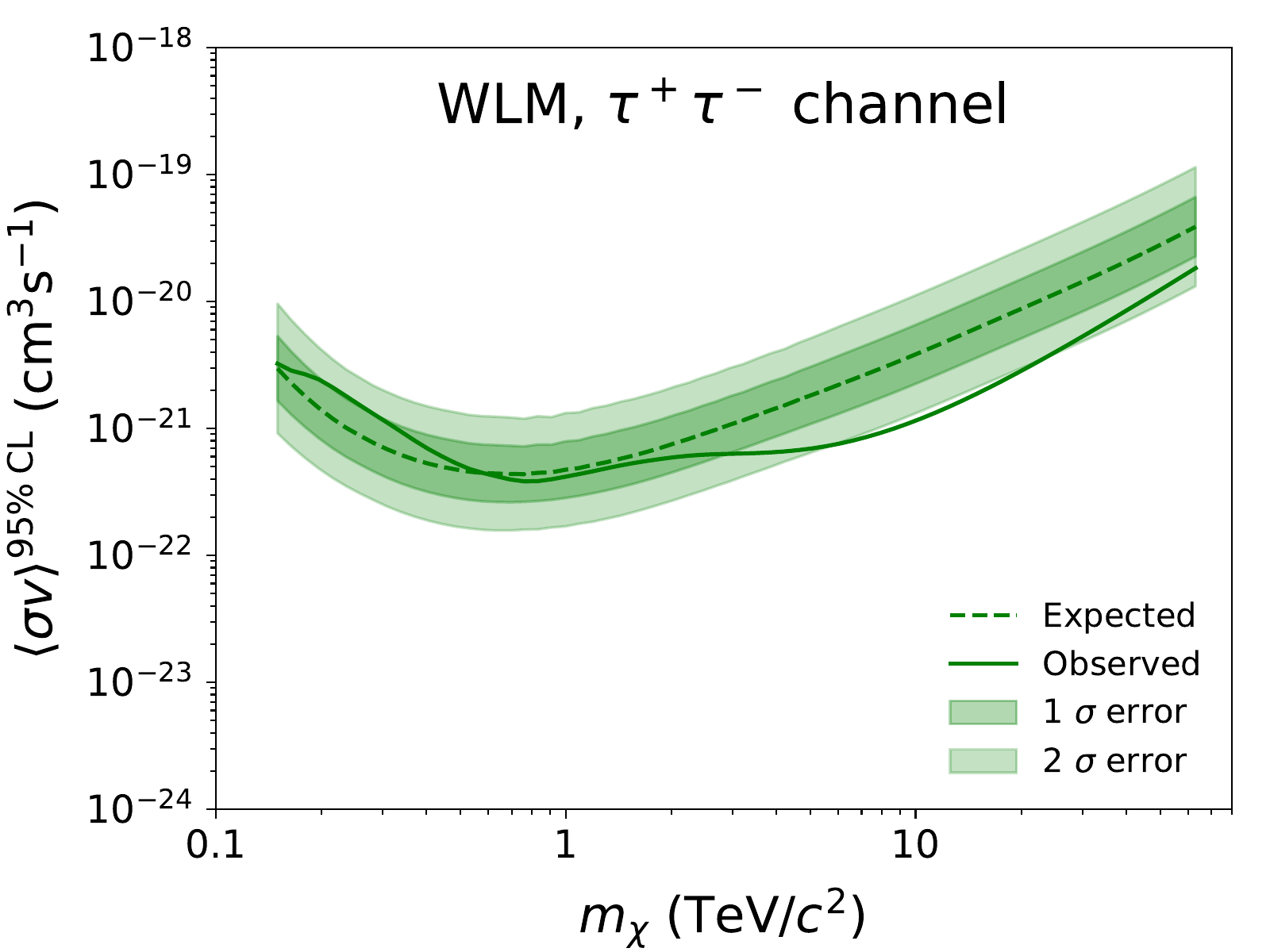}}\\
 \caption{Upper limits on the annihilation cross section $\langle \sigma v \rangle$ at 95\% CL for WLM in the $t\bar{t}$, $e^+e^-$, $\mu^+\mu^-$, and $\tau^+\tau^-$ annihilation channels. These upper limits include the uncertainties on the \J factor. The solid lines are the observed limits, the dashed lines the mean expected limits and the dark (resp. light) bands are the 1~$\sigma$ (resp. 2~$\sigma$) containment bands.}
 \label{UL_sigmav_2}
 \end{figure}

As no significant excess has been found in the region of interest, upper limits on the DM velocity-weighted annihilation cross section $\langle \sigma v \rangle$ at 95\% CL versus the DM mass are computed using the log-likelihood ratio method for the continuum channels $W^+W^-$, $Z^+Z^-$, $b\bar{b}$, $t\bar{t}$, $e^+e^-$, $\mu^+\mu^-$, $\tau^+\tau^-$, and for the mono-energetic $\gamma \gamma$ channel.
Each annihilation channel is treated individually which corresponds to a branching ratio of $B_f = 100\%$ in Eq.~(\ref{big_formula}) and all the spectra are simulated using Pythia~\cite{Cirelli:2010xx} with final state radiative corrections~\cite{Ciafaloni:2010ti} taken into account. The continuum channels and the $\gamma \gamma$ channel are treated in the same way, except that for the latter the upper limit is computed up to the highest detected energy (9.8~TeV) since the expected signal is a delta function at an energy equal to the DM mass.
As described above, we also include the uncertainties on \J as a nuisance parameter in our analysis. 
A 15\% uncertainty in the scale of the energy measurement is taken into account and added in quadrature to the total uncertainty.
The test statistic TS is computed numerically using a standard minimization algorithm.
Figures \ref{UL_sigmav_1} and \ref{UL_sigmav_2} show the 95\% CL upper limits obtained for all the annihilation channels. The mean expected limits and 1-2~$\sigma$ containment bands are derived from a sample of 300 Poisson realizations of the background events in the ON and OFF regions. The mean expected limits correspond to the mean of the distribution of $\log_{10} \langle \sigma v \rangle$ on these Poisson realizations and the uncertainty bands are given by the standard deviation of this distribution.

The observed upper limits on $\langle \sigma v \rangle$ at 95\% CL reach a few $10^{-21}$ $\text{cm}^3 \text{s}^{-1}$ for most of the continuum annihilation channels at a DM mass of 1~TeV.
For the $\tau^+ \tau^-$ annihilation channel the value of the upper limit on $\langle \sigma v \rangle$ is about $4 \times 10^{-22}$ $\text{cm}^3 \text{s}^{-1}$ for a 1~TeV DM mass. 
The $\gamma \gamma$ channel gives a more constraining result with a $\langle \sigma v \rangle$ of about $5 \times 10^{-24}$ $\text{cm}^3 \text{s}^{-1}$ at 370~GeV. %\textcolor{blue}{We note that these upper limits could have been derived using another DM density profile parametrization, preferably describing a cored DM distribution such as the Burkert profile~\cite{Burkert:1995yz} and provided that the errors on the density parameters are available. Using this profile, we obtain a relative difference of $+1.7\%$ on the \J factor within $0.12\degree$ compared to the use of the \textit{coreNFW} profile.}

\begin{figure}[ht]
 \centering{\includegraphics[scale=0.5]{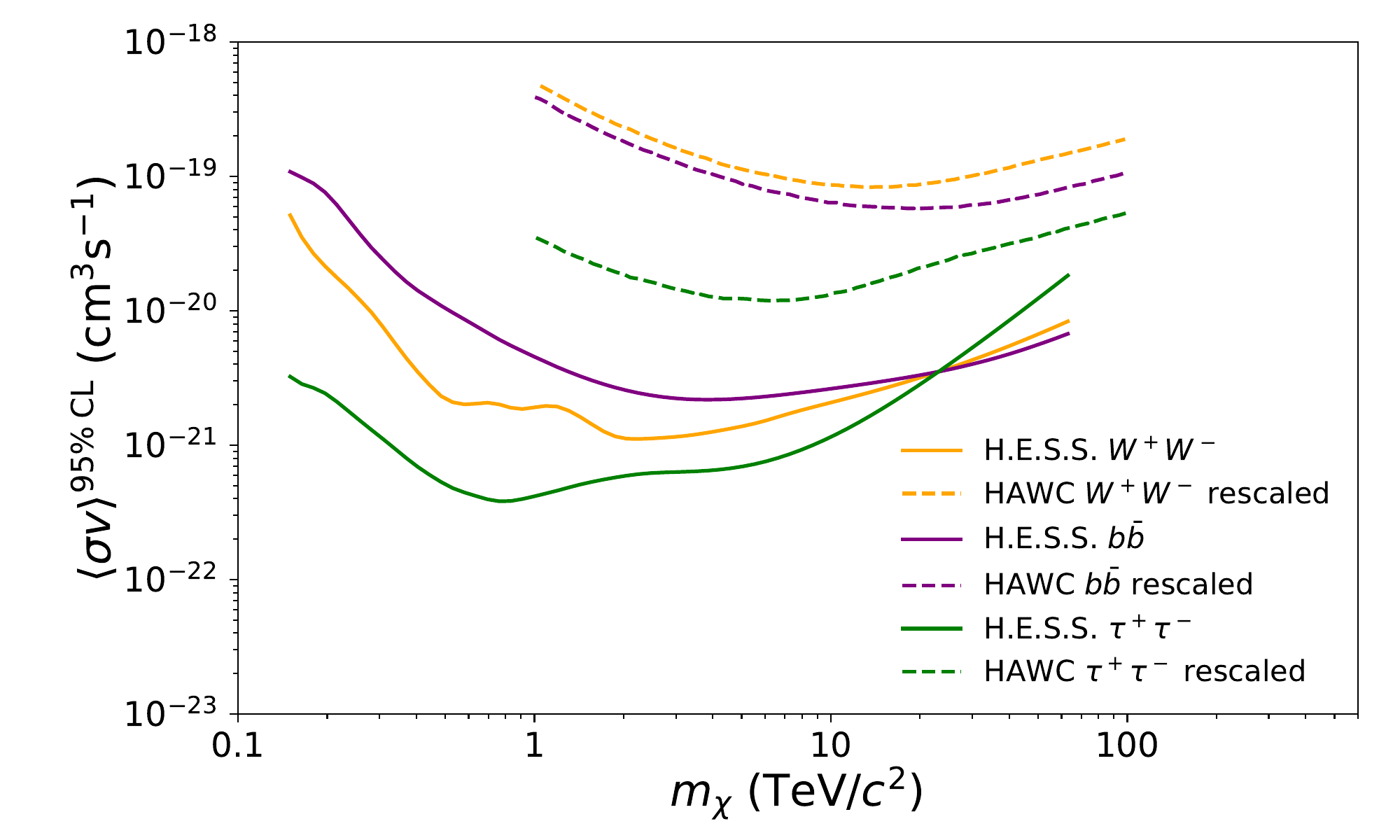}}\\
 \caption{Comparison of the upper limits for the annihilation channels $W^+W^-$, $b\bar{b}$, and $\tau^+\tau^-$ between this results and the result of the HAWC experiment for the WLM galaxy~\cite{Cadena:2017ldx}. The HAWC limits have been rescaled with the \J factor computed with the {\it coreNFW} profile used in this analysis.
}
 \label{HESS_HAWC}
 \end{figure}

The HAWC experiment~\cite{Cadena:2017ldx} showed preliminary results on 31 dwarf irregular galaxies for a total of 760 days of observation, including the study of WLM. Five annihilation channels were analysed by HAWC: $W^+W^-$, $b\bar{b}$, $t\bar{t}$, $\mu^+\mu^-$, and $\tau^+\tau^-$. The authors use for the DM halo a Burkert profile~\cite{Burkert:1995yz}, calculating the \J factor up to the virial radius. In order to compare with our results, we compute the \J factor with the {\it coreNFW} profile used here up to the virial radius, and rescale the HAWC limits accordingly. The comparison with the rescaled HAWC results is shown in Fig.~\ref{HESS_HAWC} for three annihilation channels. The H.E.S.S. results are more constraining on the whole DM mass range, by up to a factor of more than 200 depending on the annihilation channel.

The results obtained on WLM in this work can be discussed in light of recent H.E.S.S. results obtained on ultra-faint dwarf spheroidal galaxies~\cite{Abdallah:2020sas}.
The latter have a higher total $J$ factor with $\log_{10} (\J / \text{GeV}^2 \text{cm}^{-5} \rm{sr})$ from $18.7$ to $19.6$. 
The total (statistical and systematic) uncertainties can reach up to $0.9$~\cite{Bonnivard:2015tta}, where the statistical uncertainty is related to the number of measured stellar tracers. The statistical uncertainty for WLM is much smaller due to the well measured stellar and gas dynamics~\cite{R16}. No quantitative study has been performed so far on the systematic uncertainties on WLM \J factor, and therefore they are not considered in our upper limit computations. In the case of the ultra-faint dwarf spheroidal galaxies, taking into account the estimated uncertainties on the \J factor due to the limited number of tracers can degrade the upper limits by an order of magnitude~\cite{Abdallah:2020sas}.
Studying a dwarf irregular galaxy such as WLM provides a new physical DM-dominated environment to search for potential DM signals, with other types of inherent uncertainties.

%%%%%%%%%%%%%%%%%%%%%%%%%%%%%%%%%%%%%%%%%%%%%%%%%%%%%%%%%%%%%%%%%%%%%%%
\section{\label{sec:Conclusion}Conclusion}
%%%%%%%%%%%%%%%%%%%%%%%%%%%%%%%%%%%%%%%%%%%%%%%%%%%%%%%%%%%%%%%%%%%%%%%

Dwarf irregular galaxies are arguably relevant targets for the indirect search of DM through $\gamma$-ray detection. 
Their DM profile is well constrained by the gas they contain, leading to small uncertainties on the \J factor. 
With its recent 18 hour observations of WLM, one of the most promising dwarf irregular galaxies, H.E.S.S. is the first imaging atmospheric Cherenkov telescope array to observe such a galaxy to search for DM annihilation signals. The DM distribution of WLM can be well parametrized by a {\it coreNFW} profile, which allows the computation of the \J factor with a very good precision. For the whole galaxy, we obtain a value of $\log_{10} (\J_{\theta_{\rm{vir}}}  / \text{GeV}^2 \text{cm}^{-5} \rm{sr}) = 16.91 \substack{+0.10\\-0.09}$. As no detection of a significant excess signal has been made in the region of interest, upper limits on the annihilation cross section at 95\% CL have been derived for continuum annihilation channels, $W^+W^-$, $Z^+Z^-$, $b\bar{b}$, $t\bar{t}$, $e^+e^-$, $\mu^+\mu^-$, $\tau^+\tau^-$, and the prompt mono-energetic emission $\gamma \gamma$. In the case of a continuum spectrum, the most constraining limit is given by the $\tau^+\tau^-$ channel with a $\langle \sigma v \rangle$ value of about $4 \times 10^{-22}$~$\text{cm}^3 \text{s}^{-1}$ at a DM mass of 1~TeV. For the mono-energetic $\gamma\gamma$ channel, the limit on $\langle \sigma v \rangle$ reaches about $5 \times 10^{-24}$~$\text{cm}^3 \text{s}^{-1}$ at a DM mass of 370~GeV.
The upper limits derived in this work improve by a factor of at least 10 to more than 200 compared to the limits obtained by the HAWC experiment~\cite{Cadena:2017ldx}. Dwarf irregular galaxies are complementary targets to dwarf spheroidal galaxies and provide an alternative dark matter target for the next generation of Cherenkov telescopes.

%%%%%%%%%%%%%%%%%%%%%%%%%%%%%%%%%%%%%%%%%%%%%%%%%%%%%%%%%%%%%%%%%%%%%%%
\section*{Acknowledgements}
%%%%%%%%%%%%%%%%%%%%%%%%%%%%%%%%%%%%%%%%%%%%%%%%%%%%%%%%%%%%%%%%%%%%%%%

We thank Francesca Calore, LAPTh Annecy, for the useful discussions on the theoretical part of this study, as well as Justin Read, University of Surrey, for his insight about the DM distribution of WLM and for providing the input parameters of the dark matter profile used in this project.
The support of the Namibian authorities and of the University of Namibia in facilitating 
the construction and operation of H.E.S.S. is gratefully acknowledged, as is the support 
by the German Ministry for Education and Research (BMBF), the Max Planck Society, the 
German Research Foundation (DFG), the Helmholtz Association, the Alexander von Humboldt Foundation, 
the French Ministry of Higher Education, Research and Innovation, the Centre National de la 
Recherche Scientifique (CNRS/IN2P3 and CNRS/INSU), the Commissariat à l’énergie atomique 
et aux énergies alternatives (CEA), the U.K. Science and Technology Facilities Council (STFC), 
the Knut and Alice Wallenberg Foundation, the National Science Centre, Poland grant no. 2016/22/M/ST9/00382, 
the South African Department of Science and Technology and National Research Foundation, the 
University of Namibia, the National Commission on Research, Science \& Technology of Namibia (NCRST), 
the Austrian Federal Ministry of Education, Science and Research and the Austrian Science Fund (FWF), 
the Australian Research Council (ARC), the Japan Society for the Promotion of Science and by the 
University of Amsterdam. We appreciate the excellent work of the technical support staff in Berlin, 
Zeuthen, Heidelberg, Palaiseau, Paris, Saclay, Tübingen and in Namibia in the construction and 
operation of the equipment. This work benefitted from services provided by the H.E.S.S. 
Virtual Organisation, supported by the national resource providers of the EGI Federation.

\bibliographystyle{ieeetr}
\bibliography{WLM}

\end{document}